\documentclass[preprint]{elsarticle}

\usepackage{amsmath,amssymb,graphicx}
\usepackage[colorlinks=true,linktocpage=true,linkcolor=blue,citecolor=blue]{hyperref}
\usepackage{float}
\usepackage{nicefrac}
\usepackage[normalem]{ulem}
\usepackage[utf8]{inputenc}

\def\k{{\boldsymbol k}}

\def\dd{{\rm d}}

\def\ku{{k\cdot u}}
\def\trel{\tau_{\rm rel}}

\def\vsound{c_s}
\def\DK{{\rm d} K}

\newcommand{\onehalf}{{\nicefrac{1}{2}}}

\newcommand{\smallG}{{\rm \scriptscriptstyle{G}}}

\newcommand{\GZ}{{\rm \scriptscriptstyle{GZ}}}

\newcommand{\beq}{\begin{eqnarray}}
\newcommand{\eeq}{\end{eqnarray}}
\newcommand{\be}{\begin{eqnarray*}}
\newcommand{\ee}{\end{eqnarray*}}
\newcommand{\bqa}{\begin{eqnarray}}
\newcommand{\eqa}{\end{eqnarray}}

\journal{Physics Letters B}

\begin{document}

\begin{frontmatter}

\title{Bulk viscosity in a plasma of Gribov-Zwanziger gluons}

\author[a]{Wojciech Florkowski}
\ead{wojciech.florkowski@ifj.edu.pl}

\author[a]{Radoslaw Ryblewski}
\ead{radoslaw.ryblewski@ifj.edu.pl}

\author[b]{Nan Su}
\ead{nansu@th.physik.uni-frankfurt.de}

\author[c]{Konrad Tywoniuk\fnref{fn1}}
\ead{konrad@ecm.ub.edu}

\address[a]{The H. Niewodnicza\'nski Institute of Nuclear Physics, Polish Academy of Sciences, PL-31342 Krak\'ow, Poland}
\address[b]{Institut f\"ur Theoretische Physik, Goethe-Universit\"at Frankfurt am Main, D-60438 Frankfurt, Germany}
\address[c]{Deptartamento d'Estructura i Constituents de la Mat\`eria, Institut de Ci\`encies del Cosmos (ICCUB), Universitat de Barcelona, Mart\`i Franqu\`es 1, E-08028 Barcelona, Spain}

\fntext[fn1]{Presently at: Theoretical Physics Department, CERN, Geneva, Switzerland.}

\begin{abstract}
We investigate dynamic properties of a plasma whose constituents are confining gluons resulting from the Gribov-Zwanziger quantisation. In a static formulation, this system reproduces qualitatively the pure-glue equation of state and thereby encodes crucial features of the phase transition. The dynamic description proposed in this work allows us to study close-to-equilibrium transport phenomena  with the inclusion of confinement effects. In particular, we determine the non-equilibrium behaviour of the interaction measure (trace anomaly) and find the form of the bulk viscosity coefficient. The latter can be used in phenomenological applications to heavy-ion collisions.
\end{abstract}

\end{frontmatter}

\begin{keyword}
kinetic theory \sep quark-gluon plasma \sep transport coefficients \sep Gribov-Zwanziger quantisation
\PACS 05.20.Dd \sep 12.38.Aw \sep 12.38.Mh \sep 25.75.Nq
 \end{keyword}

%

\section{Introduction}
Recent experimental and theoretical studies of ultrarelativistic heavy-ion collisions suggest that the quark-gluon plasma formed in these processes behaves like a strongly-interacting and dissipative fluid~\cite{sqgp}. These observations have triggered a broad interest in the relativistic hydrodynamics of viscous systems and their respective transport coefficients \cite{hydro}. The latter, by force of the Kubo formulas, are sensitive to the long-distance dynamics of the underlying microscopic theory~\cite{Jeon:1995zm}. The expected running-coupling strength in the phenomenologically relevant regime is however on the order of unity, $g\sim{\cal O}(1)$ \cite{htl}, it is therefore not a surprise that many of the experimental features are unsatisfactorily described by conventional perturbation theory, see \cite{Haque:2014rua} for the most up-to-date results on thermodynamic quantities. Moreover, the dynamic nature of the plasma evolution severely limits the applicability of lattice methods \cite{Meyer:2011gj}. Hence the understanding of deconfinement remains a pressing challenge for heavy-ion physics.

A way to tackle these obstacles is to improve the description of soft non-Abelian gauge degrees of freedom.
In this work we introduce for the first time a non-equilibrium and dynamic description of a plasma consisting of confining gluons obtained from the Gribov-Zwanziger quantization of Yang-Mills (YM) theory \cite{Gribov:1977wm}. In this scenario, fixing the infrared (IR) residual gauge transformation in the Coulomb gauge generates a new scale that leads to an IR improved dispersion relation for gluons \cite{Gribov:1977wm},
\beq
\label{eq:GZdispersion}
E(\k) = \sqrt{\k^2 + \frac{\gamma_\smallG^4}{\k^2}} \,, 
\eeq
where $\k$ is the gluon three-momentum and $E$ the gluon energy. The Gribov parameter $\gamma_\smallG$, which governs the onset of non-Abelian effects in the IR, is found self-consistently from a gap equation \cite{Gribov:1977wm,Zwanziger:1989mf}. This leads to a suppression of soft degrees of freedom which is a necessary condition for color confinement \cite{Gribov:1977wm,Zwanziger:1989mf,Feynman:1981ss}. Thus low-momentum gluons are confining while their ultraviolet properties remain unaltered \cite{Zwanziger:1989mf}. 

The Gribov framework has been extensively studied in the vacuum, see Ref.~\cite{gribovrev} for reviews, where it is in good agreement with lattice and functional methods, especially in the Coulomb gauge~\cite{Burgio:2008jr,posviolation}. The generic modification of the IR properties of YM, embodied in Eq.~(\ref{eq:GZdispersion}), is however operative at any temperature, and the framework has also been generalized to finite temperature in order to study the deconfined phase, see e.g. \cite{Zwanziger:2004np,Zwanziger:2006sc,Fukushima:2013xsa,Su:2014rma}. In the phenomenologically relevant temperature regime the Gribov parameter $\gamma_\smallG$ is approximately a constant \cite{Fukushima:2013xsa} and, recently, its possible relation with the phase transition has also been investigated \cite{thetavac}. The resulting equation of state provides a good description of lattice data down to the vicinity of the critical temperature $T_c$ \cite{Zwanziger:2004np,Fukushima:2013xsa}. In particular, the  interaction measure, see Eq.~(\ref{eq:InteractionMeasure}), exhibits a characteristic peak near the phase transition \cite{Zwanziger:2004np}. It has also been realized that this mechanism induces long-range correlations through novel massless collective modes exclusively existing for YM theory \cite{Su:2014rma,Chernodub:2007rn}. Given these advances, it is intriguing to explore a dynamical application of the Gribov framework, especially bearing in mind the pressing challenges from heavy-ion experiments, and in the following we provide a first effort in this direction.

We will denote a hot system of gluons, governed locally by Eq.~(\ref{eq:GZdispersion}), as the Gribov-Zwanziger (GZ) plasma. Based on the discussion above, we argue that essential confinement effects are intrinsically incorporated in our setup once the Gribov dispersion relation is adopted. Since our present study focuses on the phenomenologically relevant regime, we will set $\gamma_\smallG = \text{const.}$ motivated by Refs.~\cite{Zwanziger:2004np,Fukushima:2013xsa}.

Firstly, we consider local thermodynamic properties of the GZ plasma and, secondly, determine its dynamic evolution in a boost-invariant and transversally homogeneous $(0+1)$D system \cite{Bjorken:1982qr}. This allows us to find exact solutions for in- and out-of-equilibrium evolutions using the kinetic theory approach in the relaxation time approximation \cite{Florkowski:2013lza,Florkowski:2013lya,Florkowski:2014sfa}. 

One of our main results is to determine the connection between the Gribov parameter  $\gamma_\smallG$ and the bulk viscosity $\zeta$. While the ratio of bulk viscosity to entropy density $\zeta /s$ is found to be strongly suppressed for gluons at asymptotically high temperatures \cite{htltransport}, it is expected to be enhanced in the vicinity of the YM phase transition \cite{kharzeev}, see also \cite{Sasaki:2008fg,Huang:2010sa} for results in quasi-particle models. 
This is in line with lattice results \cite{Meyer:2007dy} and a significant bulk viscosity is also obtained from calculations in strongly-coupled theories \cite{Benincasa:2005iv}. Recently the influence of bulk viscosity on observables in heavy-ion collisions \cite{Fries:2008ts,hicpheno}, as well as on the stability of the hydrodynamic evolution \cite{hydrostab}, has attracted a lot of attention. Our results highlight for the first time the direct connection of $\zeta$ to the confining properties of YM theory in the vicinity of the phase transition and its significance for phenomenological applications to heavy-ion collisions.

\section{Covariant kinetic equation setup}

In order to generalize the results \cite{Zwanziger:2004np} to a dynamical formulation, one needs to implement Lorentz covariance while respecting the dispersion relation Eq.~(\ref{eq:GZdispersion}) in the local rest frame.
The dispersion relation in Eq.~(\ref{eq:GZdispersion}) may hence be rewritten in covariant form,
\begin{eqnarray}
E(k \cdot u) = \sqrt{ (k \cdot u)^2 + \gamma_\smallG^4/(k \cdot u)^2 } \,,
\label{eq:vareps}
\end{eqnarray}
where $u$ is the four-velocity of the fluid element.\footnote{We assume that the in-medium value of $\gamma_\smallG$ is determined in the fluid element's local rest frame, where $u^\mu=(1,0,0,0)$. We introduce $k^0 \equiv |\k|$, which is the magnitude of the three-vector $\k \equiv (k_x, k_y, k_\parallel)$, and $k_\perp = (k_x^2+k_y^2)^{\onehalf}$ such that the resulting four-vector $k^{\mu} = (k^0, \k)$ has standard Lorentz transformation properties with $ k^2 = 0$. The four-vector $k$ can be interpreted as a four-momentum of a perturbative, non-interacting gluons. Equation (\ref{eq:vareps}) seems to be a straightforward generalisation of the non-covariant expression (\ref{eq:GZdispersion}), however, other version of (\ref{eq:GZdispersion}) might be also possible.} The use of Eq.~(\ref{eq:vareps}) allows us to restrict our considerations to energies that are invariably defined in the local rest frame by Eq.~(\ref{eq:GZdispersion}) where the Gribov parameter is evaluated. Further details about our covariant setup can be found in \cite{Florkowski:2015dmm}. Thus, generalizing the results of \cite{Zwanziger:2004np}, the energy density and pressure of the fluid described by the distribution function $f=f(x,k)$ are given in covariant forms by
\begin{align}
\label{eq:EnDensity}
\varepsilon &= \int \DK\,E(k \cdot u)\, f_{\rm } \,, \\
\label{eq:Pressure}
P &=  \frac{1}{3} \int \DK \,  \frac{(k\cdot u)^2}{E(k\cdot u )} \left(1 - \frac{\gamma_\smallG^4}{(k\cdot u)^4} \right) \, f \,,
\end{align}
respectively, where the integration measure is $\DK = g_0 \dd^3k (\ku)/[(2\pi)^3 k^0]$ and $g_0 = 2(N_c^2-1)$ is the degeneracy factor for gluons with $N_c$ colors ($g_0 = 16$ for SU(3)). These expressions can be used to find the interaction measure (sometimes referred to as the trace anomaly),
\begin{eqnarray}
I = \varepsilon - 3 P = 2 \int \DK \frac{\gamma_\smallG^4}{(k\cdot u)^2 \, E(k\cdot u)}  f \,.
\label{eq:InteractionMeasure}
\end{eqnarray}
This quantity is closely related to the deviation from conformality of the system under consideration. We note that $I$ vanishes identically in the $\gamma_\smallG \to 0$ limit, i.~e., for a conformal gas of massless quasi-particles.

In the case of local equilibrium, the function $f$ has the Bose-Einstein form $f_\GZ = [\exp(E(k\cdot u)/T(x)) - 1 ]^{-1}$, where the temperature $T$ can depend on space and time. The energy density and pressure obtained from Eqs.~(\ref{eq:EnDensity}) and  (\ref{eq:Pressure}) with $f=f_\GZ$ will be denoted below as $\varepsilon_\GZ$ and $P_\GZ$, respectively. The temperature dependence may in this case be eliminated to construct the equation of state of the perfect GZ plasma, $\varepsilon_\GZ = \varepsilon_\GZ(P_\GZ)$.

In order to study dynamic phenomena away from equilibrium, we employ kinetic theory in the relaxation time approximation (RTA) \cite{Florkowski:2013lya}. The $(0+1)$D system is described by the Bjorken flow vector $u^\mu = x^\mu/\tau = (t/\tau, 0,0, z/\tau)$ where $\tau = \sqrt{t^2-z^2}$ is the proper time. Furthermore, we introduce the boost-invariant variables \cite{Bialas:1987en} $v = k^0 t - k_\parallel z$ and $w = k_\parallel t -  k^0 z$, where $\ku = v/\tau = \sqrt{w^2/\tau^2 + k_\perp^2}$. Consequently, the integration measure becomes $\DK = g_0\, \dd w \, \dd^2k_{\perp}/[(2\pi)^3 \tau]$.  We note that the thermodynamic functions now depend only on the proper time, while the distribution function may depend on $\tau, w$ and $k_\perp$.

Calculating the proper time derivative of the energy density given by Eq.~(\ref{eq:EnDensity}), which in these coordinates reads $\varepsilon(\tau) = \int\DK \,E(\tau,w,k_\perp) \, f(\tau,w,k_\perp)$, one gets
\beq
\frac{\dd \varepsilon}{\dd \tau} + \frac{\varepsilon + P_\parallel}{\tau} = \int \DK \, E(\tau,w,k_\perp) \frac{\partial f(\tau,w,k_\perp)}{\partial \tau} \,,
\label{eq:deps1}
\eeq
where we identify the component of the pressure acting in the longitudinal direction as
\beq
P_\parallel = \int \DK \frac{w^2}{\tau^2 E(\tau,w,k_\perp)}\left[1-\frac{\gamma_\smallG^4}{(w^2/\tau^2 + k_\perp^2)^2} \right]f \,. 
\label{PL}
\hspace{1em}
\eeq
With $w^2/(\tau^2 E)$ replaced here by $k_\perp^2/(2 E)$  one may find the transverse pressure, 
\beq
P_\perp = \int \DK \frac{k_\perp^2}{2 E(\tau,w,k_\perp)}\left[1-\frac{\gamma_\smallG^4}{(w^2/\tau^2 + k_\perp^2)^2} \right]f \,,
\label{PT}
\hspace{1em}
\eeq
and  check that $P = (2 P_\perp + P_\parallel)/3$.\footnote{
For a $(0+1)$D system, the energy-momentum tensor is 
\begin{equation}
T^{\mu\nu} = (\varepsilon+P_\perp) u^\mu u^\nu - P_\perp g^{\mu \nu}  + (P_\parallel - P_\perp) z^\mu z^\nu,
\nonumber
\end{equation}
where $u^\mu = (t,0,0,z)/\tau$ and $z^\mu=(z,0,0,t)/\tau$. The energy density and the two pressures are given by the formulas
$\varepsilon = u_\mu u_\nu T^{\mu\nu}$, $3 P = 2 P_\perp + P_\parallel = -\Delta_{\mu\nu} T^{\mu\nu}$,
$P_\parallel = z_\mu z_\nu T^{\mu\nu}$, where $\Delta^{\mu\nu} = g^{\mu\nu} - u^\mu u^\nu$. In our case, the energy density and the pressures are expressed by the integrals over the distribution function $f$, see Eqs.~(\ref{eq:EnDensity}), (\ref{PL}), and (\ref{PT}), respectively. For more details see \cite{Florkowski:2013lza,Florkowski:2013lya,Florkowski:2014sfa} . }

The terms on the left-hand side of Eq.~(\ref{eq:deps1}) cancel due to the energy-momentum conservation,  $\partial_\mu T^{\mu\nu} =0$. This implies that the term on the right-hand side in Eq.~(\ref{eq:deps1}) should vanish as well and suggests the use of the standard RTA kinetic equation of the form
\begin{eqnarray}
\frac{\partial f(\tau,w,k_\perp)}{\partial \tau} 
= \frac{f_\GZ(\tau,w,k_\perp) - f(\tau,w,k_\perp)}{\tau_{\rm rel} (\tau)} \,,
\label{eq:ke}
\end{eqnarray}
altogether with the Landau matching condition, $\varepsilon_\GZ =\varepsilon$, given explicitly by
$\int \DK \, E(\tau,w,k_\perp) \, \left( f_\GZ  -  f \right) = 0$. We emphasise that Eq.~(\ref{eq:ke}) is applicable for the close-to-equilibrium situations but this is sufficient for derivation of our central result, namely, the formula for the bulk viscosity.\footnote{Free-streaming in Eq.~(\ref{eq:ke}) is recovered by taking the vanishing coupling limit, corresponding to simultaneously taking the relaxation time to infinity and the Gribov scale to zero, see e.g. \cite{Fukushima:2013xsa}.}

The formal solution of Eq.~(\ref{eq:ke}) is \cite{Florkowski:2013lza,Florkowski:2013lya,Florkowski:2014sfa}
\beq
f(\tau,w,k_\perp) = f_0(w,k_\perp) D(\tau,\tau_0) \label{eq:formsol} + \int_{\tau_0}^\tau \, 
\frac{d\tau^\prime}{\tau_{\rm rel}(\tau^\prime)} D(\tau,\tau^\prime)
f_\GZ(\tau^\prime,w,k_\perp),
\eeq
where the distribution function at the initial proper time $\tau_0$ is given by \newline $f(\tau_0,w,k_\perp) = f_0(w,k_\perp)$ and the damping function is given by 
\beq
D(\tau_2,\tau_1)  = \exp\left[-\int_{\tau_1}^{\tau_2} \dd\tau\, \trel^{-1}(\tau) \right] \,.
\eeq
In order to construct the solution of Eq.~(\ref{eq:formsol}) we have to know the dependence of $T$ and $\tau_{\rm rel}$ on the proper time $\tau$. Although the relaxation time $\trel$ can, in general, depend on the temperature, in this work we fix it to a constant in order to single out the genuinely novel features arising in the GZ plasma.

In the $\tau_{\rm rel} \to 0$ limit, the form of Eq.~(\ref{eq:ke}) guarantees that the actual distribution function tends rapidly to the equilibrium one and $P_\parallel = P_\perp = P_\GZ$. Thus, in order to obtain the leading-order temperature profile, we solve the well-known Bjorken hydrodynamic equation
\begin{eqnarray}
\frac{\dd \varepsilon_\GZ(T(\tau))}{\dd \tau} =   -\frac{\varepsilon_\GZ(T(\tau))+P_\GZ(T(\tau))}{\tau} \,,
\label{eq:Bjorken1}
\end{eqnarray}
which follows directly from the energy-momentum conservation \cite{Bjorken:1982qr}. In this case, the equilibrium entropy density is $s_\GZ = (\varepsilon_\GZ+P_\GZ)/T$, where $\dd\varepsilon_\GZ = T \dd s_\GZ$ and $\dd P_\GZ = s_\GZ \dd T$. The temperature dependence of the perfect GZ plasma, which we denote $T_\GZ$, follows directly from these relations, and is found by solving $\frac{\dd \ln T_\GZ(\tau)}{\dd \ln \tau} = - \,c_s^2 \left(T_\GZ(\tau) \right)$, where $c_s^2 = \partial P_\GZ/\partial \varepsilon_\GZ$ is the speed of sound of the plasma. For conformal systems, where $\varepsilon=3P$, we reproduce the well-known scaling solution $T(\tau) = T_0 (\tau_0/\tau)^{c_s^2}$ with $c_s^2 = 1/3$. In the GZ plasma, $c_s$ tend to the ideal value only in the high-$T$ limit while deviating from it in the studied temperature range.

Away from equilibrium the temperature of the system is determined from the Landau matching condition by demanding that $f$ yields the same energy density as the GZ equilibrium function $f_\GZ$. Taking the appropriate moment of the solution of the kinetic equation~(\ref{eq:formsol}) we find the integral equation
\begin{align}
\label{eq:EnDensEvolution}
\varepsilon_\GZ\left(T(\tau)\right)  &=  D(\tau,\tau_0) H_{\varepsilon}\left(\frac{\gamma_\smallG}{T(\tau_0)},\frac{\tau_0}{\tau}\right) \nonumber\\
&\hspace{1em}+ \int_{\tau_0}^\tau \frac{\dd\tau^\prime}{\tau_{\rm rel}(\tau^\prime)} D(\tau,\tau^\prime) 
H_{\varepsilon}\left(\frac{\gamma_\smallG}{T(\tau^\prime)},\frac{\tau^\prime}{\tau}\right) \,.
\end{align}
Here we have introduced the auxiliary functions
\begin{eqnarray}
H_{\varepsilon}(a,b) = \frac{g_0 \gamma_\smallG^4}{2\pi^2} \int_0^\infty \!\!\dd y \frac{y \,h_{\varepsilon}(y,b)}{ 
\exp[ a\sqrt{y^2 +1/y^2} ] -1} \,, \hspace{1em}
\end{eqnarray}
and $h_{\varepsilon}(y,b) = b \int_0^{\pi/2} \!\dd\phi  \sin(\phi) \, \beta^{-1 }\sqrt{y^4\beta^4 + 1}$ with $\beta^2(b,\phi) = b^2 \cos^2\phi + \sin^2\phi$. The temperature dependence can then implicitly be read off from the left-hand side of Eq.~(\ref{eq:EnDensEvolution}) by taking advantage of the equilibrium relation $\varepsilon_\GZ\left(T\right) = H_{\varepsilon}\left(\gamma_\smallG/T,1\right)$. The evolution of the pressure is found in a completely analogous way, i.e., by inserting the solution given in Eq.~(\ref{eq:formsol}) into the right-hand side of Eq.~(\ref{eq:Pressure}) with the temperature dependence found in the previous step.

\begin{figure}[t]
\begin{center}
\includegraphics[width=0.75\columnwidth,clip=true,trim= 0mm 0mm 3mm 0mm]{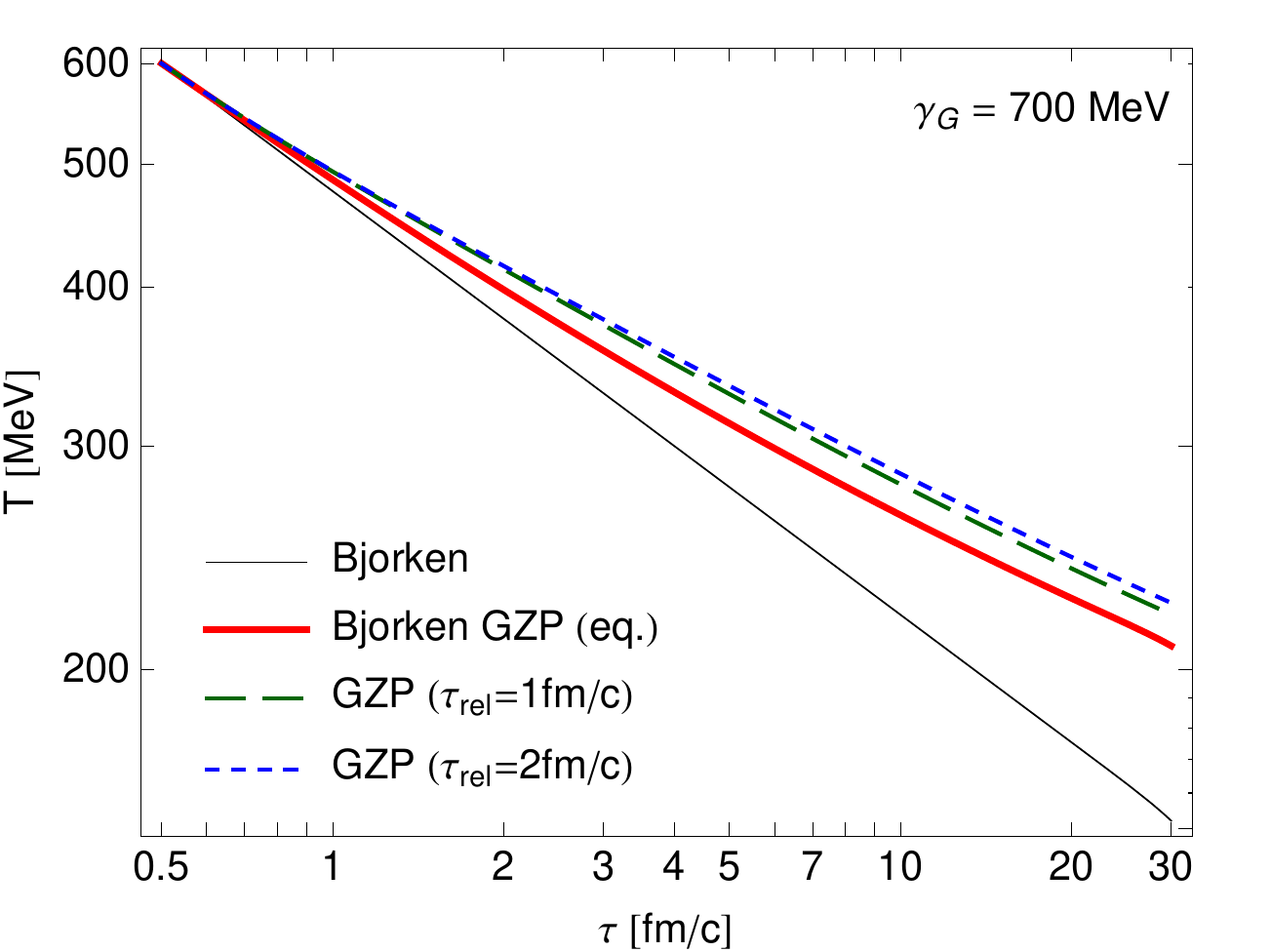}
\end{center}
\caption{Evolution of the temperature of the system. Depicted are the results for perfect fluid (black, thin line), equilibrium GZ plasma (GZP) (red, solid) and two non-equilibrium evolutions characterized by the relaxation times $\trel =$1 fm/c (dashed, green curve) and $\trel =$ 2 fm/c (dotted, blue curve) }%
\label{fig:temperature}
\end{figure}
%
\section{Thermodynamics-like quantities}
We proceed with numerical calculations based on Eq.~(\ref{eq:formsol}), where the energy density and pressure are read off from Eqs.~(\ref{eq:EnDensity}) and (\ref{eq:Pressure}), respectively. For the initial condition we choose the GZ plasma in equilibrium at a given temperature. As stated in the discussion above, we fix the Gribov parameter to the value $\gamma_\smallG =$ 700 MeV in order to obtain the best description of lattice data at $T\gtrsim T_c$ in agreement with previous studies \cite{Zwanziger:2004np}. Furthermore, we choose the initial time $\tau_0$~=~0.5~fm/c, the initial temperature $T(\tau_0)$~=~600~MeV, and study the non-equilibrium evolutions for two fixed relaxation times, $\tau_{\rm rel}$~=~1~fm/c and  $\tau_{\rm rel}$~=~2~fm/c.

The proper time dependence of the temperature is shown in Fig.~\ref{fig:temperature}. The GZ plasma (GZP) in equilibrium (solid, red curve) cools at a slower rate than the ideal gas (black, thin curve) owing to the reduction of the speed of sound in the vicinity of the critical temperature. The rate of cooling for local equilibrium is
only mildly restrained when going away from equilibrium, see the dashed, green and dotted, blue curves in Fig.~\ref{fig:temperature}  which correspond to the two equilibration times described above, respectively (we keep the same color coding in all subsequent figures). 

\begin{figure}[t]
\begin{center}
\includegraphics[width=0.75\columnwidth,clip=true,trim= 5mm 2mm 5mm 0mm]{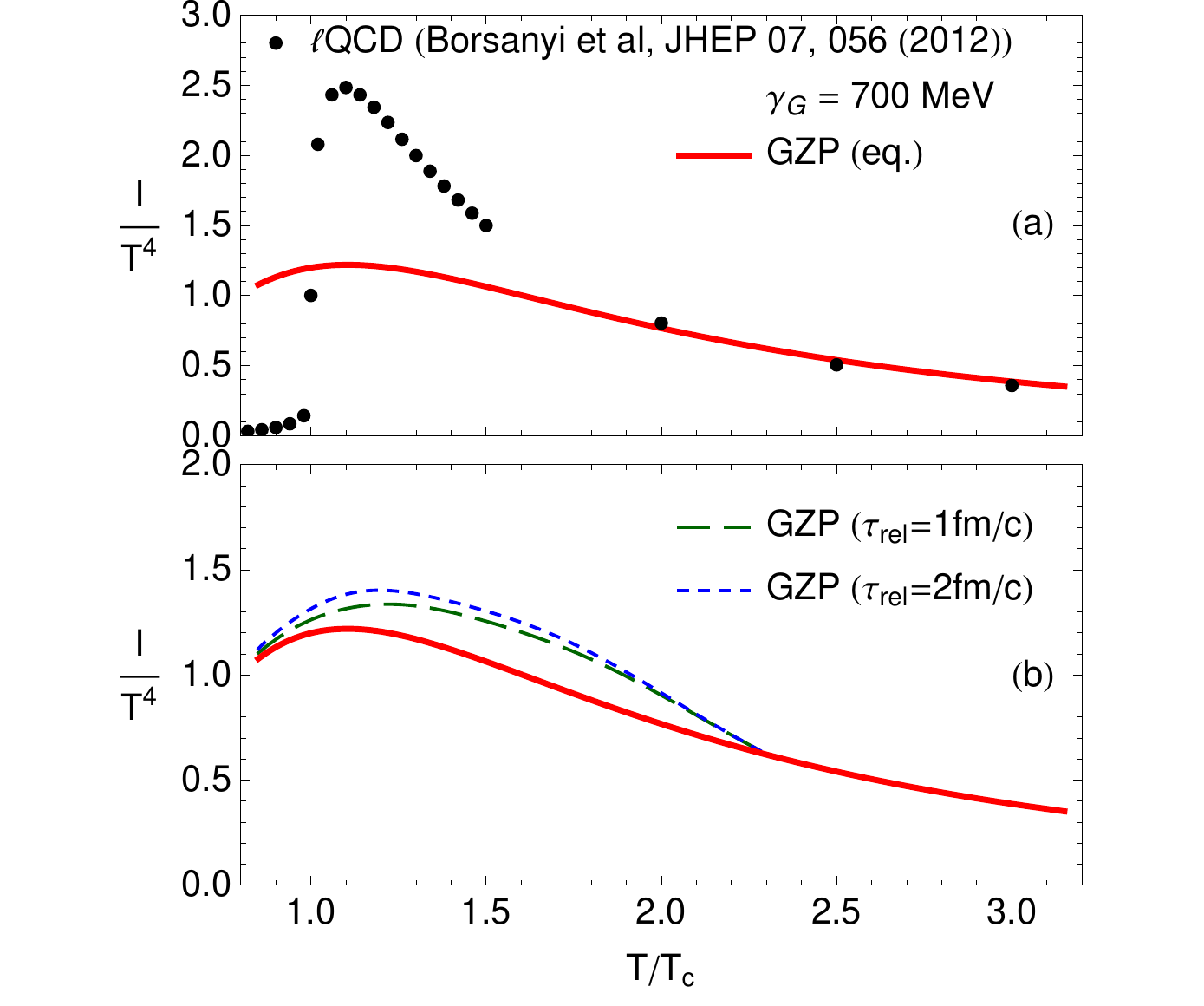}
\end{center}
\caption{Temperature dependence of the interaction measure $I$ scaled by $T^4$. Upper panel: The result for the GZ plasma (GZP) in local equilibrium (solid, red curve) together with the lattice data~\cite{Borsanyi:2012ve} (black dots). Lower panel: The two non-equilibrium results for $I$, characterized by the relaxation times $\trel =$1 fm/c (dashed, green curve) and $\trel =$ 2 fm/c (dotted, blue curve), compared to the equilibrium calculation (solid, red curve).}%
\label{fig:anomaly}
\end{figure}
The upper panel of Fig.~\ref{fig:anomaly} shows our results for the interaction measure, Eq.~(\ref{eq:InteractionMeasure}), compared to the lattice data~\cite{Borsanyi:2012ve}, where we have chosen $T_c = $ 260 MeV. The solid red line depicts the equilibrium result, which qualitatively describes the lattice data: it yields half of the peak of the anomaly in the region of the phase transition and gives a good description of the anomaly in the temperature range just above the phase transition, $T \sim 1.5-4 \,T_c$. In this way we reproduce the result of Ref.~\cite{Zwanziger:2004np}. Further improvements can only be achieved by taking into account quantum effects \cite{Fukushima:2013xsa}. The dashed and dotted lines in the lower panel of Fig.~\ref{fig:anomaly} depict the results of the non-equilibrium calculations for the two chosen values of $\trel$ and approach the solid, equilibrium line at lower temperatures, reflecting that for late times $\tau \gg \tau_\text{rel}$ the system approaches equilibrium. The results presented in the lower panel of Fig.~\ref{fig:anomaly} indicate that the trace anomaly becomes larger if the system evolves out of equilibrium. In addition, the width of the peak becomes wider. Similarly to the equilibrium results, we expect that our kinetic results are the most legitimate in the same temperature region.

\section{Bulk viscosity from the kinetic equation}
We note that the interaction measure vanishes exactly at all temperatures for a conformal theory. In our case, it is the presence of an IR non-Abelian scale in the dispersion relation Eq.~(\ref{eq:GZdispersion}), which allows us to qualitatively reproduce the YM theory around the phase transition. Although the energy density is kept equal to the equilibrium one due to the Landau matching condition, the pressure can deviate. The difference between the actual and equilibrium pressures describes the bulk viscous pressure $\Pi = P - P_\GZ$. Owing to the definition of the trace anomaly in Eq.~(\ref{eq:InteractionMeasure}) and the Landau matching, one finds
\begin{eqnarray}
\Pi = -\frac{2}{3} \int \DK \frac{\gamma_\smallG^4}{(k\cdot u)^2 \,E(k\cdot u)} 
\left( f - f_\GZ\right) \,,
\label{eq:Pi1}
\end{eqnarray}
which simply corresponds to $(I_\GZ-I)/3$. The knowledge of $\Pi$ allows us to determine the effective bulk viscosity $\zeta_{\rm eff}$ of the system through the identification
\begin{eqnarray}
\Pi = -\zeta_{\rm eff}\, \partial_\mu u^\mu =  -\zeta_{\rm eff}/\tau \,,
\label{eq:Pi2}
\end{eqnarray}
where the latter equality holds for a $(0+1)$D system. 

For small deviations from equilibrium, the effective viscosity $\zeta_{\rm eff}$ approaches the standard bulk viscosity coefficient $\zeta$. To determine $\zeta$ we seek the solution of the kinetic equation~(\ref{eq:ke}) in the form $f \approx f_\GZ + \delta f$. In the linear approximation, we find $\delta f = - \trel \dd f_\GZ/\dd\tau$ and substitute $\delta f$ into the right-hand side of Eq.~(\ref{eq:Pi1}). Using the equilibrium relation for the speed of sound, we find
\beq
\label{eq:ZetaRestFrame}
\zeta = \frac{g_0\gamma_\smallG^4 }{3\pi^2} \frac{\tau_{\rm rel}}{T}  \int_0^\infty \!\!\dd y \left[ \vsound^2 - \frac{1}{3}\frac{y^4 - \gamma_\smallG^4}{y^4+ \gamma_\smallG^4} \right] f_\GZ(1+f_\GZ),
\eeq
where $f_\GZ = \big[\exp(\sqrt{y^2+\gamma_\smallG^4/y^2}/T)-1 \big]^{-1}$,
which is one of the main results of our paper. The bulk viscosity is proportional to the relaxation time by construction. It vanishes when $\gamma_\smallG \to 0$, which is the case for a conformal, massless gas. Our setup is hence qualitatively different from the outset due to the presence of the Gribov parameter.
\begin{figure}[t]
\begin{center}
\includegraphics[width=0.85\textwidth,clip=true,trim= 5mm 5mm 0mm 5mm]{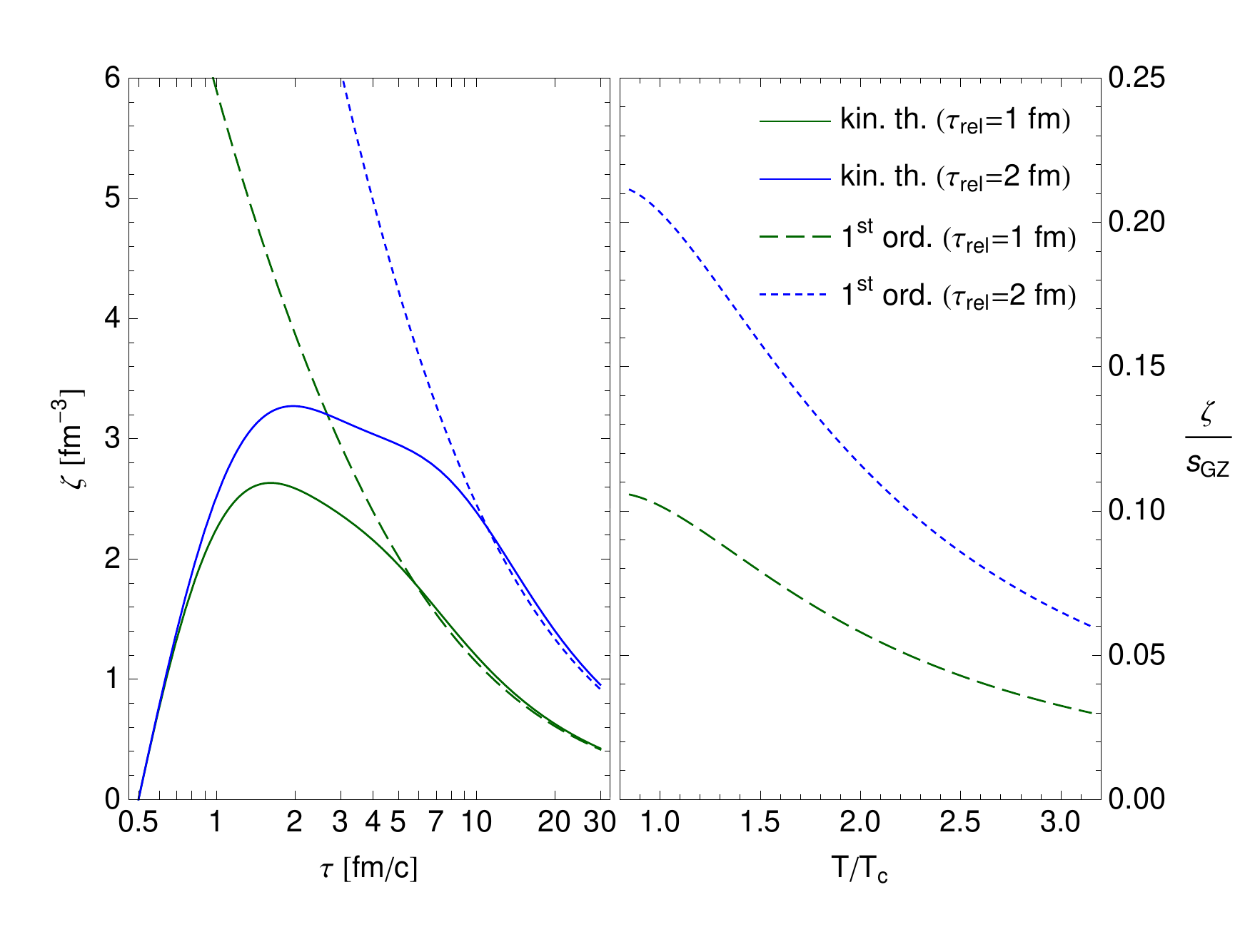}
\end{center}
\caption{Left panel: The effective bulk viscosity evaluated from the kinetic theory, Eqs.~(\ref{eq:Pi1}) and (\ref{eq:Pi2}) (solid curves), and the bulk viscosity coefficient obtained from the linearized expression in Eq.~(\ref{eq:ZetaRestFrame}) (dashed and dotted curves)). 
Right panel: Temperature dependence of the bulk viscosity coefficient scaled by the equilibrium entropy density for two values of the relaxation time.
The color coding is the same as in Figs. \ref{fig:temperature} and \ref{fig:anomaly}. 
}%
\label{fig:bulk}
\end{figure}

In the left panel of Fig.~\ref{fig:bulk} we show the proper time dependence of the effective bulk viscosity $\zeta_{\rm eff}$ obtained directly from the numerical solution of the kinetic equation using Eqs.~(\ref{eq:Pi1}) and (\ref{eq:Pi2}), and of the bulk viscosity coefficient $\zeta$ obtained from  Eq.~(\ref{eq:ZetaRestFrame}). As the system approaches equilibrium we find a good agreement between $\zeta_{\rm eff}$ and $\zeta$,  which supports the validity of Eq.~(\ref{eq:ZetaRestFrame}). 
Moreover, it is worth noting the close relationship between $\zeta$ and the deviation from conformality, understood here as $c_s^2-1/3$, see Eq.~(\ref{eq:ZetaRestFrame}). 
On a qualitative level, this is also observed in quasi-particle models \cite{Huang:2010sa}, however the magnitude of $\zeta/s_\GZ$ is significantly smaller in this case. Further improvements are necessary for the GZ gas in order to facilitate more qualitative studies.
 In a parallel work \cite{Florkowski:2015dmm}, we have derived a scaling of the ratio of bulk to shear viscosities to the speed of sound, $\zeta/\eta = \kappa_\GZ(1/3-c_s^2)$ at high temperature that is similar to the results from calculations in holographic, strongly-coupled theories \cite{Benincasa:2005iv}.
 
In the right panel of Fig.~\ref{fig:bulk}, on the other hand, we show the ratio $\zeta/s_\GZ$ as a function of temperature. Similarly to Fig.~\ref{fig:temperature}, we observe a significant increase of the bulk viscosity in the vicinity of the phase transition. We note that large bulk viscosity may be related to prolonged deviations from equilibrium \cite{Fries:2008ts}.

\section{Summary} 
In this work, we have established a novel framework for dealing with a non-equilibrium plasma of confining gluons in close to equilibrium situations. Our method is based on the improved dispersion relation for gluons, Eq.~(\ref{eq:GZdispersion}), which accounts for residual long-range correlations in the system in the deconfined phase. If local equilibrium is assumed, our system expands and cools as expected from the Bjorken model, see the red curve in Fig.~\ref{fig:temperature}. In non-equilibrium situations, our numerical results show an increase of the trace anomaly at a given effective temperature, see Fig.~\ref{fig:anomaly}. In the latter case, the system evolution is affected by dissipative phenomena, in particular, by the presence of bulk viscosity. The full and linearized solutions, derived in Eq.~(\ref{eq:ZetaRestFrame}), for the bulk viscosity agree at large times and indicate a rise of the ratio $\zeta/s$ close to the phase transition as expected in a strongly coupled plasma. 

The features above provide an improved understanding of the hot plasma produced in heavy-ion collisions, in line with expectations from lattice data, and implies that bulk viscosity should be implemented in their hydrodynamic modeling. Finally, we note that the framework proposed herein can be naturally extended to include other transport phenomena such as, for example, the shear viscosity, and be systematically improved in order to account for a more involved and realistic form of the relaxation time and a temperature dependence of the Gribov parameter \cite{Zwanziger:2006sc,Fukushima:2013xsa}. We leave these developments, which would allow for more qualitative assessments of the dynamical properties of the GZ plasma, for future works.

\section*{Acknowledgments}
Research supported in part by Polish National Science Center grants No. DEC-2012/05/B/ST2/02528, No. DEC-2012/06/A/ST2/00390 (W.F.) and No. DEC-2012/07/D/ST2/02125 (R.R.).
K.T. would like to thank the H. Niewodnicza\'nski  Institute of Nuclear Physics, where this work was initiated, for hospitality. 
K.T. was supported by a Juan de la Cierva fellowship and by the Spanish MINECO under projects  FPA2013-46570 and 2014SGR104, partially by MDM-2014-0369 of ICCUB (Unidad de Excelencia 'Mar\'ia de Maeztu'), by the Consolider CPAN project and by FEDER. This research project has been supported by a Marie Sklodowska-Curie Individual Fellowship of the European Commission's Horizon 2020 Programme under contract number 655279 ``ResolvedJetsHIC''.


\section*{References}
\begin{thebibliography}{99}


\bibitem{sqgp}
  M.~Gyulassy and L.~McLerran,
  Nucl.\ Phys.\ A {\bf 750} (2005) 30;
  E.~V.~Shuryak,
  Nucl.\ Phys.\ A {\bf 750} (2005) 64.


\bibitem{hydro}
  P.~Romatschke,
  Class.\ Quant.\ Grav.\  {\bf 27} (2010) 025006;
  S.~Jeon and U.~Heinz,
  arXiv:1503.03931 [hep-ph].
  

\bibitem{Jeon:1995zm} 
  S.~Jeon and L.~G.~Yaffe,
  Phys.\ Rev.\ D {\bf 53} (1996) 5799.


\bibitem{htl}
  J.~O.~Andersen and M.~Strickland,
  Annals Phys.\ {\bf 317} (2005) 281;
  N.~Su,
  Commun.\ Theor.\ Phys. {\bf 57} (2012) 409;
  N.~Su,
  Int.\ J.\ Mod.\ Phys.\ A {\bf 30} (2015) 1530025.

\bibitem{Haque:2014rua}
  N.~Haque, A.~Bandyopadhyay, J.~O.~Andersen, M.~G.~Mustafa, M.~Strickland and N.~Su,
  JHEP {\bf 1405} (2014) 027.



\bibitem{Meyer:2011gj} 
  H.~B.~Meyer,
  Eur.\ Phys.\ J.\ A {\bf 47} (2011) 86.


\bibitem{Gribov:1977wm} V.\ Gribov, Nucl.\ Phys.\ B {\bf 139} (1978) 1.

\bibitem{Zwanziger:1989mf}  D.~Zwanziger,  Nucl.\ Phys.\ B {\bf 323} (1989) 513.

\bibitem{Feynman:1981ss} 
  R.~P.~Feynman,
  Nucl.\ Phys.\ B {\bf 188} (1981) 479.

\bibitem{gribovrev}
  Y.~L.~Dokshitzer and D.~E.~Kharzeev,
  Ann.\ Rev.\ Nucl.\ Part.\ Sci.\  {\bf 54} (2004) 487;
  N.~Vandersickel and D.~Zwanziger,
  Phys.\ Rept.\  {\bf 520} (2012) 175.

\bibitem{Burgio:2008jr} 
  G.~Burgio, M.~Quandt and H.~Reinhardt,
  Phys.\ Rev.\ Lett.\  {\bf 102} (2009) 032002;
  Phys.\ Rev.\ D {\bf 86} (2012) 045029.


\bibitem{posviolation} 
 R.~Alkofer and L.~von Smekal,
  Phys.\ Rept.\  {\bf 353} (2001) 281;
   A.~Maas,
  Phys.\ Rept.\  {\bf 524} (2013) 203.

\bibitem{Zwanziger:2004np} D.~Zwanziger, Phys.\ Rev.\ Lett.\ {\bf 94} (2005) 182301.

\bibitem{Zwanziger:2006sc}  D.~Zwanziger, Phys.\ Rev.\ D {\bf 76} (2007) 125014. 

\bibitem{Fukushima:2013xsa} K.~Fukushima, N.~Su, Phys.\ Rev.\ D {\bf 88} (2013) 076008.

\bibitem{Su:2014rma} 
  N.~Su and K.~Tywoniuk,
  Phys.\ Rev.\ Lett.\  {\bf 114} (2015) 161601.

\bibitem{thetavac} D.~E.~Kharzeev, E.~M.~Levin, Phys.\ Rev.\ Lett {\bf 114} (2015) 242001.
 
\bibitem{Chernodub:2007rn} 
  M.~N.~Chernodub and V.~I.~Zakharov,
  Phys.\ Rev.\ Lett.\  {\bf 100} (2008) 222001.
 
 
 \bibitem{Bjorken:1982qr} 
  J.~D.~Bjorken,
  Phys.\ Rev.\ D {\bf 27} (1983) 140.


\bibitem{Florkowski:2013lza} 
  W.~Florkowski, R.~Ryblewski and M.~Strickland,
  Nucl.\ Phys.\ A {\bf 916} (2013) 249.
  
\bibitem{Florkowski:2013lya} 
  W.~Florkowski, R.~Ryblewski and M.~Strickland,
  Phys.\ Rev.\ C {\bf 88} (2013) 024903.
  
  \bibitem{Florkowski:2014sfa} 
  W.~Florkowski, E.~Maksymiuk, R.~Ryblewski and M.~Strickland,
  Phys.\ Rev.\ C {\bf 89} (2014) 054908.
  
\bibitem{htltransport}
  P.~B.~Arnold, C.~Dogan and G.~D.~Moore,
  Phys.\ Rev.\ D {\bf 74} (2006) 085021;
  G.~D.~Moore and O.~Saremi,
  JHEP {\bf 0809} (2008) 015.
  
\bibitem{kharzeev}
  D.~Kharzeev and K.~Tuchin,
  JHEP {\bf 0809} (2008) 093;
  F.~Karsch, D.~Kharzeev and K.~Tuchin,
  Phys.\ Lett.\ B {\bf 663} (2008) 217.
  
\bibitem{Sasaki:2008fg} 
  C.~Sasaki and K.~Redlich,
  Phys.\ Rev.\ C {\bf 79} (2009) 055207.

\bibitem{Huang:2010sa} 
  X.~G.~Huang, T.~Kodama, T.~Koide and D.~H.~Rischke,
  Phys.\ Rev.\ C {\bf 83} (2011) 024906.
  
   \bibitem{Meyer:2007dy} 
  H.~B.~Meyer,
  Phys.\ Rev.\ Lett.\  {\bf 100} (2008) 162001.

\bibitem{Benincasa:2005iv} 
  P.~Benincasa, A.~Buchel and A.~O.~Starinets,
  Nucl.\ Phys.\ B {\bf 733} (2006) 160.


\bibitem{Fries:2008ts} 
  R.~J.~Fries, B.~Muller and A.~Schafer,
  Phys.\ Rev.\ C {\bf 78} (2008) 034913.
  
\bibitem{hicpheno}
  P.~Bozek,
  Phys.\ Rev.\ C {\bf 81} (2010) 034909;  
  A.~Monnai and T.~Hirano,
  Phys.\ Rev.\ C {\bf 80} (2009) 054906;
  G.~S.~Denicol, T.~Kodama, T.~Koide and P.~Mota,
  Phys.\ Rev.\ C {\bf 80} (2009) 064901;
  J.~Noronha-Hostler, G.~S.~Denicol, J.~Noronha, R.~P.~G.~Andrade and F.~Grassi,
  Phys.\ Rev.\ C {\bf 88} (2013) 044916;
  S.~Ryu, J.-F.~Paquet, C.~Shen, G.~S.~Denicol, B.~Schenke, S.~Jeon and C.~Gale,
  arXiv:1502.01675 [nucl-th].

\bibitem{hydrostab}
  G.~Torrieri and I.~Mishustin,
  Phys.\ Rev.\ C {\bf 78} (2008) 021901;
  K.~Rajagopal and N.~Tripuraneni,
  JHEP {\bf 1003} (2010) 018.
  
  
\bibitem{Florkowski:2015dmm}
  W.~Florkowski, R.~Ryblewski, N.~Su and K.~Tywoniuk,
  arXiv:1509.01242 [hep-ph].
  
  

  \bibitem{Bialas:1987en} 
  A.~Bialas, W.~Czyz, A.~Dyrek and W.~Florkowski,
  Nucl.\ Phys.\ B {\bf 296} (1988) 611.


\bibitem{Borsanyi:2012ve} 
  S.~Borsanyi, G.~Endrodi, Z.~Fodor, S.~D.~Katz and K.~K.~Szabo,
  JHEP {\bf 1207} (2012) 056.


\end{thebibliography}
\end{document}